\documentstyle [prb,aps,epsfig,multicol]{revtex}
\begin{document}                                                    
\draft                                                      
\title{Many-Polaron Effects in the Holstein Model }
\author{Sanjoy Datta, Arnab Das, and Sudhakar Yarlagadda}
\address{ Saha Institute of Nuclear Physics, Calcutta, India}
\date{\today}
\maketitle

\begin{abstract}
We derive an effective polaronic interaction Hamiltonian,
{\it exact to second order in perturbation}, for the spinless
one-dimensional Holstein model. 
 The small parameter is given by the ratio of the hopping term ($t$)
to the polaronic energy ($g^2 \omega _0$)
in all the region of validity for our perturbation;
however, the exception being the regime
of extreme anti-adiabaticity ($t/\omega_0 \le 0.1$)
and small electron-phonon coupling ($g < 1$) where 
the small parameter is $t/\omega_0$. 
We map our polaronic Hamiltonian onto a 
next-to-nearest-neighbor interaction
anisotropic Heisenberg spin model. By studying the
mass gap and the power-law exponent of the
spin-spin correlation function for our Heisenberg spin model,
we analyze the Luttinger
liquid to charge-density-wave transition at half-filling in
the effective polaronic Hamiltonian.
We calculate the structure factor at all fillings and find
that the spin-spin correlation length decreases as one deviates from
half-filling. We also extend our derivation of polaronic Hamiltonian
to $d$-dimensions.
\end{abstract}

\nopagebreak
\section{INTRODUCTION}
Understanding the many-body aspects of the Holstein model \cite{holstein}
has been a long standing open problem. 
Significant progress has been made in understanding
the single polaron problem
through analytic treatments
in the small and large polaron limits
and numerical approaches in the intermediate size case \cite{alex}.
While studies of the many-polaron problem, involving spin degrees of freedom,
yielded interesting insights for bipolarons and their phase transitions
\cite{alex,alex2},
the simpler case of spinless many-polaron problem has received
only scant attention \cite{alex3}.
With the renewed interest in strongly correlated 
manganite  systems \cite{cnr}
(which are of high electronic density
when the electron-phonon
interactions are supposed to be important)
it is imperative that the effective interaction between 
polarons be understood so that 
a serious attempt at explaining the
rich phase diagram of these systems be made.
Although studying manganites
demands knowledge of the effective Jahn-Teller polaronic interaction, 
 in the low-doped case the effective Hamiltonian at $0$ K
 for the occupied states can be taken as
a Holstein model \cite{dattays}. 
Thus a good understanding, involving exact results,
 of the simpler effective polaronic interaction
 for the Holstein model,
which has been elusive so far, is highly desirable.
Furthermore, an effective polaronic Hamiltonian even in the simplest
spinless 1D case would also
be quite useful for modelling Luttinger liquid (LL) to 
charge-density-wave (CDW)
transitions in half-filled systems.
Quasi-1D 
organic conjugated polymers  [such as ${\rm  (CH)_{x}}$], 
charge transfer salts [such as TTF(TCNQ)], and inorganic blue bronzes 
(e.g., ${\rm K_{0.3}MoO_3}$)
\cite{yamaji} are good candidates for such broken symmetry
in the ground state leading to unit cell doubling.

The present paper, using Lang-Firsov transformation \cite{lang},
provides a transparent perturbative approach to deriving
 the {\it effective Hamiltonian of
interacting polarons in d-dimensions}
when the band narrowing is significant.
The resulting polaronic Hamiltonian, {\it exact to second order in
 perturbation}, is studied in 1D
at $0$ K for density-density correlation
effects. The correlation function
exponent of the concomitant quasi-long range order
is demonstrated to be useful in characterizing the Luttinger liquid 
and the charge-density-wave phases of the system.

\section{EFFECTIVE POLARONIC HAMILTONIAN}
We begin in 1D by taking
the unperturbed Hamiltonian to be the non-interacting
 polaronic term\cite{ys}
\begin{eqnarray}
H_0 =&& 
 \omega_0 \sum_{j} a^{\dagger}_{j} a_{j} -
 \omega_0 g^2 \sum_{j} c^{\dagger}_{j} c_{j} 
\nonumber \\
&& - J_1 \sum_{j} ( c^{\dagger}_{j} c_{j+1} + {\rm H.c.}) , 
\end{eqnarray}
with the perturbation being
\begin{eqnarray}
H^{\prime} = \sum_{j}H_j =
 - J_1 \sum_{j} ( c^{\dagger}_{j} c_{j+1}
\{
{\cal{S}}_{+}^{j \dagger} {\cal{S}}^{j}_{-}
-1 \} + {\rm H.c.}) , 
\end{eqnarray}
where $H_0 + H^{\prime}$ make up the Lang-Firsov transformed
 Holstein Hamiltonian
 and $c_j$ ($a_j$) is the fermionic (phononic) destruction
operator,
$\omega_0$ the Debye frequency,
 ${\cal{S}}^{j}_{\pm} = \exp[\pm g( a_{j} - a_{j+1})]$,
$J_1 = t \exp(- g^2)$,
with $t$ being the hopping term, 
 and $g^2 \omega_0$ the polaronic binding energy.
The eigen states are given by $|n,m \rangle \equiv |n \rangle_{el} \otimes
 |m\rangle_{ph}$
with $|0,0\rangle$ being the ground state with 
zero phonons. The eigen energies
are $E_{n,m}=E_{n}^{el}+E_{m}^{ph}$. 
Since   $\langle 0,n^{\prime} |H^{\prime}|n,0\rangle = 0$, the
 first order perturbation term is zero
and the relevant excited states correspond to states with non-zero phonons.
Next, we represent $|m\rangle_{ph}$ in real space
 as phononic excitations at different sites
with one phonon state being $a^{\dagger}_{j}|0\rangle_{ph}$ which is
 $N$-fold degenerate and can correspond to any site $j$.
On the other hand, the electronic state $|n\rangle_{el}$ is represented in
the momentum space. We will now calculate second-order perturbation 
terms {\it exactly}
\begin{eqnarray*}
E^{(2)} = \sum_{l,j}
\sum_{n,m} \frac{\langle 0,0| H_l |n,m\rangle \langle m,n| H_j |0,0\rangle}
{E_{0,0}-E_{n,m}} .
 \end{eqnarray*}                               
Now $E_{n}^{el} - E_{0}^{el} \sim J_1$ and 
 $ \Delta E_m \equiv E_{m}^{ph} - E_{0}^{ph}$ is
 a non-zero integral multiple of $\omega_0$.
Assuming 
 $ J_1 << \omega_0$ (which certainly is true for realistic
values of $2 < t/\omega_0 < 4$ and $6 < g^2 < 10$ found in manganites) 
 and using $\sum_n |n\rangle \langle n|=I$ we get
 the corresponding second-order term in the effective Hamiltonian
 for polarons to be
\begin{eqnarray*}
H^{(2)} = \sum_{l,j}
\sum_{m} \frac{\langle 0|_{ph} H_l |m\rangle_{ph}
 \langle m|_{ph} H_j |0\rangle_{ph}}
{-\Delta E_m} .
 \end{eqnarray*}                               
In the above equation, the term $H_j$ produces phonons at sites $j$ and $j+1$.
Hence  to match that its counterpart $H_l$ should produce phonons in at least
one of the two sites $j$ and $j+1$. Thus the index 
$l=j-1,$ $j,$ or $j+1$.
Next on defining $P_{\pm}(j;m) \equiv
 \langle 0|_{ph} {\cal{S}}^{j}_{\pm} -1 |m\rangle_{ph}$
and $b_j  \equiv c^{\dagger}_j c_{j+1}$ we get
\begin{eqnarray}
H^{(2)} = -  \sum_{j,m}&&
\frac{J_1^2}{\Delta E_m}[\{
b^{\dagger}_j b_j P_{+}(j;m) 
+b_{j-1} b_j P_{-}(j-1;m) 
\nonumber \\
&& 
+b_{j+1} b_j P_{-}(j+1;m)\}P_{+}^{\dagger}(j;m)
\nonumber \\
&& +
\{b_j b^{\dagger}_j P_{-}(j;m) 
+b^{\dagger}_{j-1} b^{\dagger}_j P_{+}(j-1;m) 
\nonumber \\
&& 
+b^{\dagger}_{j+1} b^{\dagger}_j P_{+}(j+1;m)\}P_{-}^{\dagger}(j;m)
] .
 \end{eqnarray}                               
Then using $(a^{\dagger})^n |0\rangle = \sqrt{n!}|n^{\prime}\rangle$
with $|n^{\prime}\rangle$ being a state with $n$ phonons
we get the effective polaronic Hamiltonian to be
\begin{eqnarray}
H^{pol}_{eff} = && -g^2 \omega_0 \sum_j n_j
 - J_1 \sum_j (c^{\dagger}_{j}c_{j+1} + {\rm H.c.}) 
\nonumber \\
&& + J^z \sum_j n_j n_{j+1}
+2 J_2 \sum_j  (c^{\dagger}_{j-1}n_j c_{j+1} + {\rm H.c.} )
\nonumber \\
&&
- J_2 \sum_j  (c^{\dagger}_{j-1} c_{j+1} + {\rm H.c.} )
- J^z \sum _j n_j  ,
\label{Heff}
 \end{eqnarray}                               
where 
$J^z \equiv \frac{J_1^2}{\omega _0}[4 f_1(g) + 2 f_2(g)]$,
and 
$J_2 \equiv \frac{J_1^2}{\omega _0} f_1(g)$
with $f_1 (g) \equiv \sum_{n=1}^{\infty} \frac {g^{2n}}{n! n}$
and  $f_2 (g) \equiv 
\sum_{n=1}^{\infty}
\sum_{m=1}^{\infty}
 \frac {g^{2(n+m)}}{n!m!(n+m)}$.
It is of interest to note that the {\it single} polaronic energy
part of the above Hamiltonian matches with the 
self-energy expression at $k=0$ obtained by  Marsiglio \cite{mars} and
the self-energy at a  general
$k$ by Stephan \cite{stephan}
and lends credibility to our results.
Furthermore, in the work of
 Hirsch and Fradkin \cite{fradkin},
the coefficient of nearest-neighbor interaction 
agrees with our coefficient $J^z$
 for large values of $g$
 while the coefficients of the next-to-nearest-neighbor (NNN)
 hopping are in disagreement with those of ours and
the results of Refs. \cite{mars,stephan}.
Next we make the connection that, on using Wigner-Jordan transformation
$\sigma ^{+}_{i}=\Pi_{j<i} (1-2 n_j)c^{\dagger}_i$,
we can map the effective polaronic Hamiltonian exactly
(up to a constant) on to the following
NNN anisotropic Heisenberg spin chain:
\begin{eqnarray}
H_{eff}^{spin} = && -g^2 \omega_0 \sum_j \sigma^{z}_j
 -J_1 \sum_j (\sigma^{+}_{j}\sigma^{-}_{j+1} + {\rm H.c.}) 
\nonumber \\
&& + J^z \sum_j 
\sigma^{z}_j \sigma^{z}_{j+1} 
-J_2 \sum_j 
(\sigma^{+}_{j-1} \sigma^{-}_{j+1} + {\rm H.c.} )  ,
\label{Hspin}
 \end{eqnarray}                               
where the coefficient of the first term represents coupling to a longitudinal
magnetic field.
Although the 
NNN interactions in the above Hamiltonian
do not produce frustration, nevertheless the Hamiltonian
cannot be solved by coordinate Bethe ansatz \cite{ibose}.
Hence we take
recourse to analyzing the properties of the effective Hamiltonian 
numerically by using modified Lanczos technique
(see Ref. \cite {gagli} for details).  

\section{LUTTINGER LIQUID TO CDW TRANSITION}
The spin Hamiltonian $H^{spin}_{eff}$
in Eq. (\ref{Hspin})
with $J_2 = 0$,  i.e., without NNN
 interaction,
has been shown to undergo a LL to
CDW state transition at zero magnetization
($\sum_j \sigma^{z}_j = 0$) when $J^z =2 J_1$ and 
at non-zero magnetizations is always a LL \cite{haldane}.
On including a non-zero $J_2$, the disordering effect increases because
the NNN
interaction is only in the transverse
 direction and the LL to CDW
transition will occur at higher values of $J^z$.
We expect that including $J_2$ does not change the universality class
and that the central charge $c=1$.

We first study the static spin-spin correlation function on rings with even  
number of sites $N$ 
 and extract information about the critical exponent.
The static spin-spin correlation function 
for a chain of length $N$ is given by
$W_l(N)=\frac{4}{N} \sum_{i=1}^{N} \langle S_i^z S_{i+l}^z\rangle $
and has the asymptotic behavior
$\lim_{N \rightarrow \infty } W_l(N)
 \approx \frac{A (-1)^{l}} {l^{\eta}} $
 for the anisotropic Heisenberg model
when $l \gg 1$ \cite{luther}.
 Furthermore, $A$ is an unknown constant and 
$1 < \eta \leq 2$
when the system is in disordered (or LL) state, $\eta =1$
is the transition point to antiferromagnetic (CDW) state, 
and $\eta=0$ means system is totally antiferromagnetic (or CDW).

We will now derive an analytic expression for the critical exponent $\eta$
based on the work of Luther and Peschel \cite{luther}.
The effective polaronic Hamiltonian given by
Eq. (\ref{Heff})  can be written in momentum space as 
\begin{eqnarray}
H^{pol}_{eff}&=&-2J_1\sum_{k}\cos(k)c^{\dagger}_{k}c_{k}
+\frac{J^z}{N}\sum_{q}\cos(q)\rho(q)\rho(-q)
\nonumber\\
 & & \mbox{}-\frac{4 J_2}{N}\sum_{k,k',q}\cos(k+k')
c^{\dagger}_{k+q}c^{\dagger}_{k^{\prime}}c_{k^{\prime}+q}c_{k}
\nonumber\\
& & -2J_2
\sum_{k}\cos(2k)c^{\dagger}_{k}c_{k} ,
\end{eqnarray}
where $\rho(q)=\sum_{p}c^{\dagger}_{q+p}c_{p}$. 
Furthermore, constant terms have been ignored.
Next, we linearize the kinetic energy term close to the
 Fermi points and 
 follow it up with the bosonization procedure. Then on taking exchange effects
 into account, as pointed out by Fowler\cite{fowler}, we obtain the following
 bosonized Hamiltonian 
\begin{eqnarray}
H^{pol}_{bos}&=&\left[\frac{4\pi J_1+4J^z+8J_2}{N}\right]
\sum_{k>0;i=1,2}
\rho_{i}(k)\rho_{i}(-k)
\nonumber\\
   & &\mbox{}+\left[\frac{8J^z-32 J_2}{N}\right]
\sum_{k>0}\rho_{1}(k)\rho_{2}(-k) .
\label{Hbos}
\end{eqnarray}
It is important to note that only the forward scattering part involving
 the coefficient $J^z$ contributes to the self energy correction.

Now, to calculate the critical exponent $\eta$, we will follow 
the usual procedure and diagonalize the bosonized Hamiltonian of 
Eq. (\ref{Hbos}) using
 the following transformations,
\begin{eqnarray*}
\rho_{1}(k)&=&\bar{\rho}_{1}(k)\cosh\phi+\bar{\rho}_{2}(k)\sinh\phi ,
\end{eqnarray*}
 and 
\begin{eqnarray*}
\rho_{2}(k)&=&\bar{\rho}_{2}(k)\cosh\phi+\bar{\rho}_{1}(k)\sinh\phi .
\end{eqnarray*}
Then, on setting the coefficient of the off-diagonal term equal to zero
in the transformed Hamiltonian, we get
\begin{equation}
\tanh2\phi=-\frac{2J^z-8J_2}{2 \pi J_1 +2 J^z+ 4 J_2} .
\label{tanh}
\end{equation}
Using Eq.(\ref{tanh}) we obtain 
\begin{equation}
\eta=2e^{2\phi}=2{\left[\frac{1+\frac{6J_2}
{\pi J_1}}{1+\frac{2J^z}{\pi J_1}-\frac{2 J_2}
{\pi J_1}}\right]}^{\frac{1}{2}} .
\end{equation}

\section{RESULTS AND DISCUSSION}
It is known that for an anisotropic Heisenberg spin chain,
 when $N/2$ is even the correlation function goes to zero 
smoothly as the longitudinal interaction goes to zero \cite{lieb}.
Hence we have calculated $W_{N/2}(N)$ only for odd values
of $N/2$ with $N$ = 6, 10, 14, 18, and 22
at $J_1$ = 1 and different values of $J^z$ and $J_2$.
Using a linear least squares fit for a plot of  $\log W_{N/2}(N)$
versus $\log N$ we obtained 
 the value of $\eta$ from the slope at each value of $J^z$ and
$J_2$. The error in $\eta$ for all curves is within $\approx 0.05$
and hence verifies
that one has the expected quasi-long range order.
 For $J^z = 2$ and $J_2 = 0$ we get $\eta = 0.96 \pm 0.05$
which includes the exact value of $\eta = 1$ obtained from
Bethe ansatz. Thus we expect the $\eta$ values obtained by
our procedure to be reasonably accurate. Since for $J^z > 2$ and $J_2 = 0$
we obtain a CDW state, by increasing $J_2$ at any $J^z > 2$
 we should increase the disordering
effect  and hence we see in Fig. 1 that $\eta$ value increases. We find that
for $J^z \approx 6$ the value of $\eta$ becomes slightly negative but
 with magnitude
within the error of 0.05. At higher $J^z$ values ($\approx 10$ 
 and higher) $\eta$ tends
to zero. At small values of $J^z$ ($ \le 0.5 $) as $J_2$ increases initially
$\eta$ increases even to values above 2 and then decreases to values below 2.
We think that this interesting feature is due to smaller values
of $J_2$ enhancing the disordering effect while the larger
values of $J_2$ increase correlations build up
with the system becoming less LL like.
However the behavior at $J^z =0$ and $J_2 > 0$
needs further understanding.
Our derived analytic expression, reliable at small values of $J_2/J_1$
and $\eta < 2$, shows that $\eta$ does increase with increasing values
of $J_2$ and gives values reasonably close to the numerical ones for
$J^z / J_1 < 1$.

We will now consider the mass gap, at the half-filled
state for the Hamiltonian $H=H^{pol}_{eff}+g^2\omega_0 \sum_j n_j$
 [see Eq. (\ref{Heff})],
defined as twice the energy difference between the
ground state with $1+N/2$ electrons and the ground state of the
 $N/2$ electronic system. 
The mass gap is calculated for rings with
$N$ = 10, 12, 14, 16, 18, and 20 sites.
Including $N$ = 6 and 8 only increases the error.
 The mass gap plot in Fig. 2 is obtained 
using finite-size scaling by
plotting mass gap versus $1/N$ and extrapolating the linear least
square fit to the value
corresponding to $1/N =0$.
 In the plot the size of the symbol is larger than the error.
 From the inset of the plot of mass gap versus $J^z$
at various values of $J_2$, we see that the mass gap
goes to zero at $J^z \approx 1.4$ at $J_2 =0$ which is a significant
underestimation of the transition value of $J^z =2 $. Also on comparing
with Fig. 1, we again notice that the LL to CDW transition
value of $J^z$ at different $J_2$ values is 
grossly underestimated. Furthermore, as expected, mass gap increases
(decreases) monotonically with $J^z$ ($J_2$) at a fixed $J_2$ ($J^z$).

We will now discuss the region of validity
for our model, given by Eq. (\ref{Heff}) and as
depicted in Fig. 3 (region above the lower curve),
in the two-dimensional parameter space of $g$ and $t/\omega_0$.
Firstly, since we use the assumption that $\omega_0 \gg J_1$
in our derivation, we choose the validity condition as 
 $\omega_0 \geq 10 J_1$. Next, we would like
the second order energy term $E^{(2)}$ in the perturbation series
 to be much smaller
than the unperturbed term $E_{0,0}$.
We find that for
$t/\omega_0 \leq 1$, the condition $\omega_0 = 10 J_1$ 
produces a boundary on which
the ratio $E_{0,0}/E^{(2)} > 5$ 
with the ratio increasing rapidly as $t/\omega_0 $ decreases. 
As for $t /\omega_0 \geq 2$, we find that the condition 
$g^2 \omega_0 \geq 3 J^z$ is more restrictive than the first 
one ($\omega_0 \geq 10 J_1$) and produces a ratio of 
$E_{0,0}/E^{(2)} > 3$ at the boundary.
Next we will discuss the LL to CDW phase transition boundary obtained from
$\eta = 1$ condition and  depicted by the upper curve in Fig. 3.
We find that the phase transition points 
 lie within the
region of validity only for $t/\omega_0\leq 0.6$.
In the region to the right of the dashed vertical line and below the
region-of-validity curve the phase boundary
cannot be determined using our model.
It is important to note that the {\it experimentally realistic parameter
regime $6 < g^2 < 10$ and $2 < t/\omega_0 < 4$ lies mostly inside the region
of validity}.
Upon comparing our numerical phase transition results with those
of Bursill {\it et al.}  \cite{hamer}, we find that for small values
of $t/\omega_0 \leq 0.1$ the critical $g_c$ values agree well.
However  for larger values the results do not agree.
At $t /\omega_0 = 0.5$  our
$g_c = 1.45 \pm 0.02$ (with  
$E_{0,0}/E^{(2)} > 17$ and $\omega/J_1 > 15$)
 is noticeably smaller
than the $g_c = 1.63(1)$ of Ref. \cite{hamer} and at $t/\omega_0 = 1$
we find that $g_c < 1.52$ whereas Bursill {\it et al.} get $g_c = 1.61(1)$.
As for $t/\omega_0 \geq 2$, our region of validity lies above the phase
transition boundary given in Ref. \cite{hamer}.
However, interestingly, the  numerical estimates of the critical $g_c$ 
by Hirsch and Fradkin \cite{fradkin} are consistent with our
results with their $g_c$ value agreeing with ours at $t/\omega_0 = 0.5$, while
at higher values of $t/\omega_0$ their $g_c$ values lie outside our 
region of validity.

Now that the region of validity has been identified,
we will analyze within this region the small parameter for our
pertubation theory.
For $g > 1$,   
 one approximates $f_1 (g) \sim e^{g^2}/g^2 $
and $[2 f_1(g) + f_2 (g)] \sim e^{2 g^2} / {2 g^2}$
with the approximations becoming
exact in the limit $g \rightarrow \infty$. Then the effective polaronic
Hamiltonian of 
Eq. (\ref{Heff}), for the case $g > 1$, simplifies to
\begin{eqnarray}
H^{pol}_{eff} \sim  -g^2 \omega_0 [&&   \sum_j n_j
 + \zeta e^{-g^2} \sum_j (c^{\dagger}_{j}c_{j+1} + {\rm H.c.}) 
\nonumber \\
&&
+ \zeta ^2 e^{-g^2} \sum_j  \{ c^{\dagger}_{j-1} (1 -2 n_j ) c_{j+1}
 + {\rm H.c.} \}
\nonumber \\
&&
 + \zeta ^2 \sum_j n_j (1 -  n_{j+1})
 ] ,
\label{Heff_sp}
 \end{eqnarray}                               
where $\zeta \equiv t/g^2 \omega_0$ is the polaron size parameter.
In the region of validity for our model, when the adiabaticity
parameter $t/\omega_0 > 0.2 $,
we have the constraints $g > 1$ and $g^2 \omega _0 \ge 2t$.
Thus we see that, for  
the region $t/\omega_0 > 0.2 $,
the polaron size parameter
$\zeta $ is the small parameter.

Now, for the extreme anti-adiabatic regime of $t/\omega_0 \le 0.1$,
all values of $g$ are allowed by our model. 
When $g > 1$, again Eq. (\ref{Heff_sp}) is valid with the same small
parameter $\zeta$. However, when $g < 1$, we make the approximations 
 $f_1 (g) \sim g^2$ and 
 $[2 f_1(g) + f_2 (g)] \sim 2 f_1 (g) $
with the approximations becoming exact in the limit $g \rightarrow 0$.
Then, for $g < 1$, 
the effective polaronic
Hamiltonian given by 
Eq. (\ref{Heff}) becomes
\begin{eqnarray}
H^{pol}_{eff} \sim  -g^2 \omega_0 [ && \sum_j n_j
 + \zeta e^{-g^2} \sum_j (c^{\dagger}_{j}c_{j+1} + {\rm H.c.}) 
\nonumber \\
&&
+ 
 \left ( {\frac{t}{\omega_0}} \right ) ^2 e^{-2 g^2} 
 \sum_j  \{ c^{\dagger}_{j-1} (1 -2 n_j ) c_{j+1} + {\rm H.c.} \} 
\nonumber \\
&&
 +
4 \left ( {\frac{t}{\omega_0}} \right ) ^2 e^{-2 g^2} 
 \sum_j n_j (1 -  n_{j+1})
] .
\label{Heff_sp2}
 \end{eqnarray}                               
 Thus, for the regime $t/\omega_0 \le 0.1$ 
and $g < 1$, the adiabaticity parameter $t/\omega_0$
is the small parameter in Eq. (\ref{Heff_sp2})
with $g \sim 1$ corresponding to small polarons and
 $g \rightarrow 0$ (such that $g^2 \omega_0 << t$)   
corresponding to large polarons.

Finally, we also study the static structure factor
$S_N (k) \equiv \sum_{l=1}^{N} e^{ikl} W_l(N)$ .
The structure factor offers information about the correlation
lengths even in LL phase at all filling factors $n$.
In fact, the correlation length decreases with increasing
 width of the structure
factor near its peak at $2 \pi n$.
We first observe that $\sum _k S_N (k) = N$
 and that $S_N (0) = 4N(n-0.5)^2$
all of which are borne out by both the plots in Fig. 4
done at $N = 20$.
The plots are only for $k = 2 \pi m /N$ with $m = 0, 1, ..., N/2$
as $S_N (k)$ is symmetric about $\pi$ and are
only for $n \leq 0.5$ because of particle-hole symmetry.
Fig. 4(b), plotted for 
$t/\omega_0 = 0.5$ and $g = g_c = 1.45 $ (or $J^z = 2.53$ and $J_2 = 0.245$),
 corresponds to LL-CDW transition point at $n = 0.5$, 
 while Fig. 4(a), done for 
the realistic values of $t /\omega_0 = 3$ and $g =  3 $
 (or $J^z \approx 3000$ and $J_2 \approx 0.38$),
 depicts situation deep inside the CDW phase at $n = 0.5$.
Now, we know that at $n = 0.5$, the structure factor
 $S_N (\pi ) \sim \int \frac{d l}{l ^{\eta}}$ and hence diverges
for $\eta \leq 1$ (CDW regime) as $N \rightarrow \infty$ with
the divergence being faster as we go deeper inside the CDW regime.
 Also the structure factor remains finite at $2 \pi n$
for all other filling factors even when
$N \rightarrow \infty$ because here the system is always a LL.
From plot (a) we infer that deep inside the CDW state $S_N( \pi ) \approx N$
and $S_N( k \neq \pi ) \approx 0$. As for the CDW transition point
depicted in plot (b) at $n = 0.5$,
the structure factor peaks sharply but more gradually at
$k = \pi$.
In both plots the largest peak occurs for $n = 0.5 $ with peak
size diminishing and curve width increasing as values
of $n$ decrease. Thus, we see that for $n \neq 0.5$ also,
short range correlations exist with correlation length
decreasing as deviation from half-filling increases.

Lastly and importantly, using arguments similar to the 1D
 case,
we have also derived the effective polaronic Hamiltonian in $d$-dimensions
to be \cite{sdys}
\begin{eqnarray}
H^{pol}_{eff} = && -g^2 \omega_0 \sum_j n_j
 - J_1 \sum_{j, \delta} c^{\dagger}_{j}c_{j+\delta}  
\nonumber \\
&&
- J_2 \sum_{j, \delta , \delta ^{\prime} \neq \delta}
  c^{\dagger}_{j+\delta ^{\prime}}(1 -2 n_j ) c_{j+ \delta} 
\nonumber \\
&&
- 0.5 J^z \sum _{j, \delta} n_j (1 -  
  n_{j+ \delta})
,
\label{Heff2}
 \end{eqnarray}                               
where $\delta$ corresponds to nearest neighbor \cite{alex4}.

In conclusion, we have derived an effective polaronic Hamiltonian for
 the spinless 1D Holstein model which is found to be valid in most of
the experimentally realistic regime.
 We mapped the effective Hamiltonian
onto a next-to-nearest-neighbor anisotropic Heisenberg Hamiltonian.
Using modified Lanczos technique extensively, we computed
the static spin-spin correlation exponent $\eta$  and the mass gap at
half-filling for general
values of the parameters in the effective spin Hamiltonian.
The mass gap values were found to significantly underestimate
the critical electron-phonon coupling $g_c$. In contrast,
the $\eta$ values were found to give reliable estimates of $g_c$ and 
consequently were used
to determine the LL-CDW quantum phase transition.
The structure factor calculations revealed that correlation length
diminishes with increasing deviation from half filling.
Lastly, our approach, {\it exact to second order in perturbation},
is extended to obtain 
an effective polaronic Hamiltonian
in $d$-dimensions also.
Our work opens up a whole host of future challenges
such as: (1) Extension to finite
 temperatures and studying 
metal-insulator transition; 
(2) Including Hubbard interaction $U$;
(3) Analyzing the $d$-dimensional model in Eq. (\ref{Heff2})
\cite {sdys}; and (4) Deriving analogous effective Hamiltonian for
  Jahn-Teller systems.

\begin{figure}
\resizebox*{6.8in}{5.0in}{\rotatebox{0}{\includegraphics{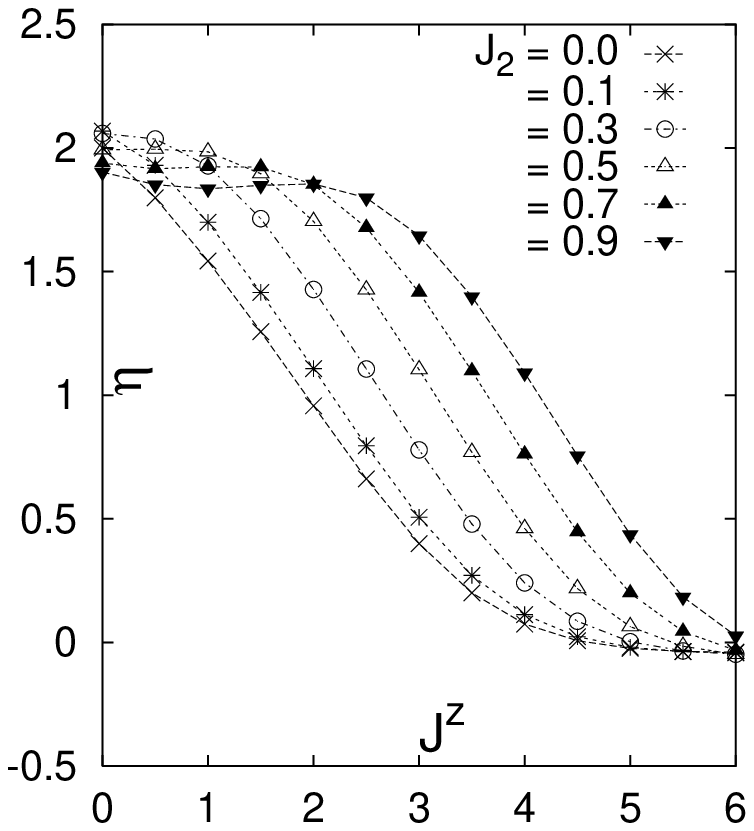}}}
\vspace*{0.5cm}
\caption[]{Plot of spin-spin correlation exponent for various values
of $J^z$ and $J_2$. The dashed lines are guides to the eye.}
\label{scaling1}
\end{figure}
\begin{figure}
\resizebox*{6.2in}{5.0in}{\rotatebox{0}{\includegraphics{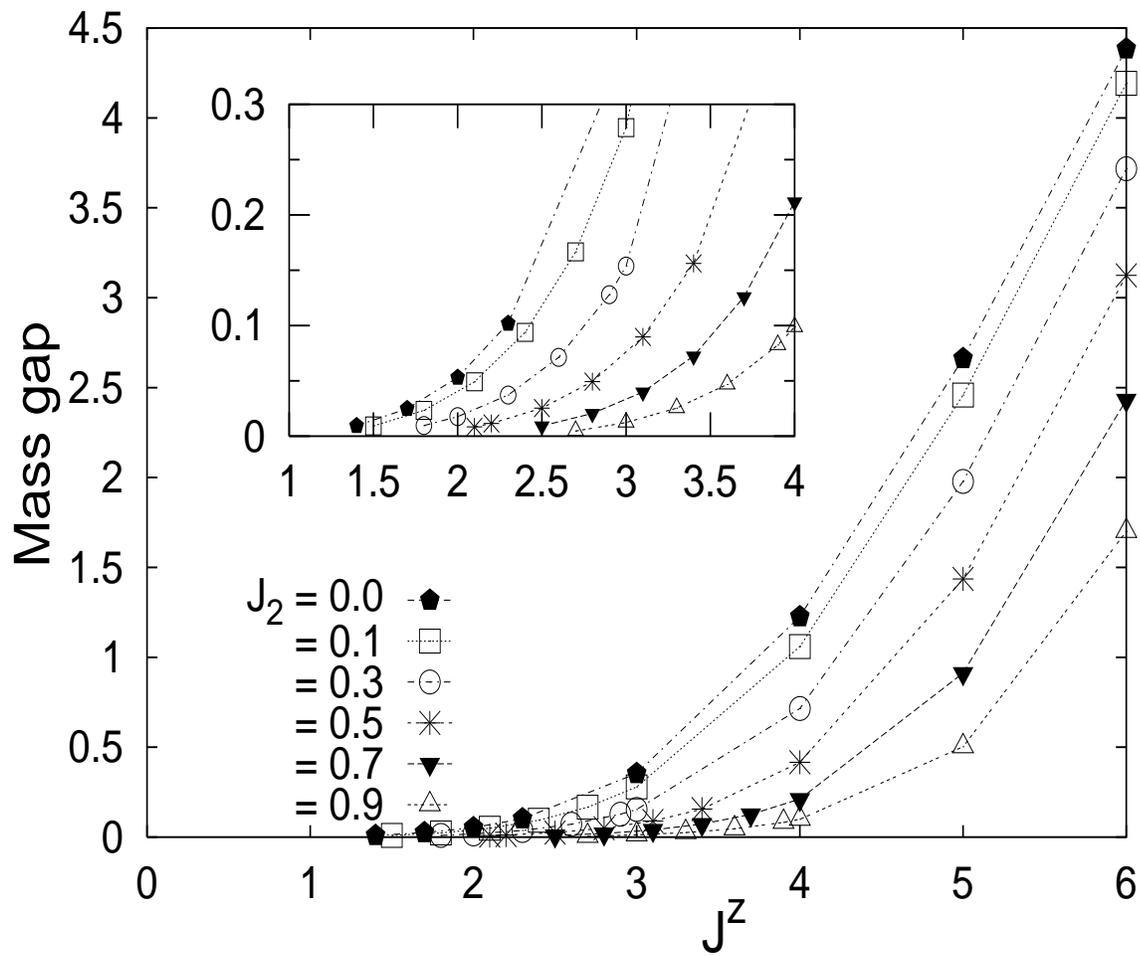}}}
\vspace*{0.5cm}
\caption[]{Mass gap dependence on $J^z$ and $J_2$.}
\label{scaling2}
\end{figure}
\begin{figure}
\resizebox*{6.2in}{4.0in}{\rotatebox{0}{\includegraphics{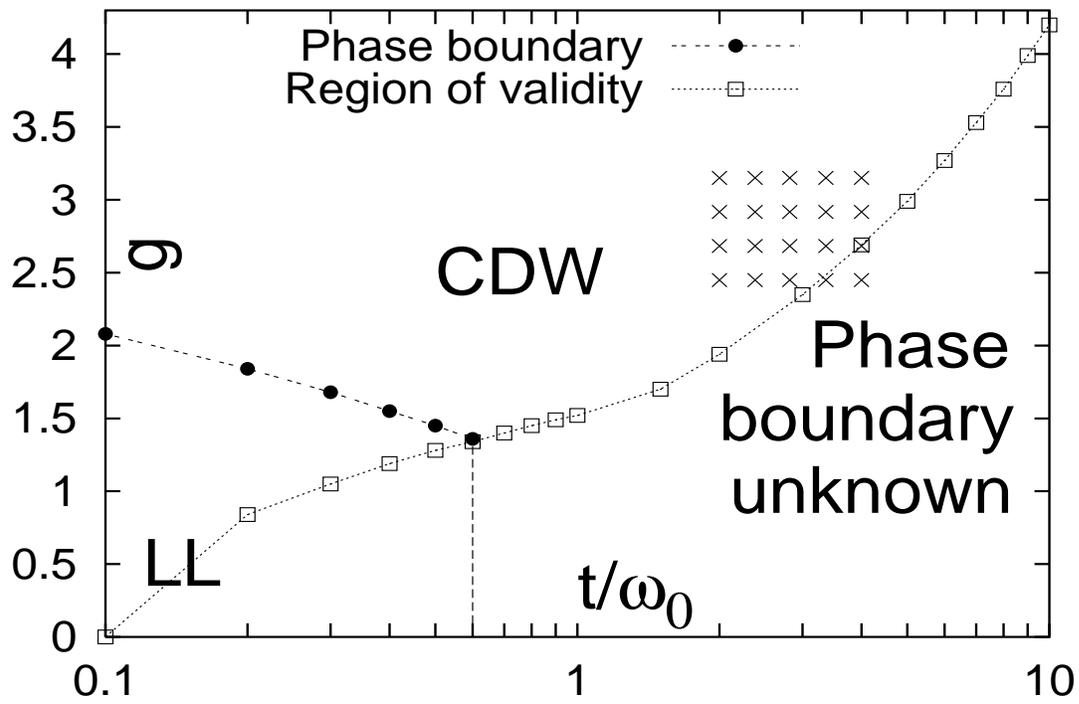}}}
\vspace*{0.5cm}
\caption[]{Plot of region of validity boundary and LL-CDW
phase boundary for 
$g$ versus $t/\omega_0$. The errors are smaller than the symbols.
The crosses depict realistic regime.}
\label{scaling3}
\end{figure}
\begin{figure}
\resizebox*{6.8in}{5.0in}{\rotatebox{0}{\includegraphics{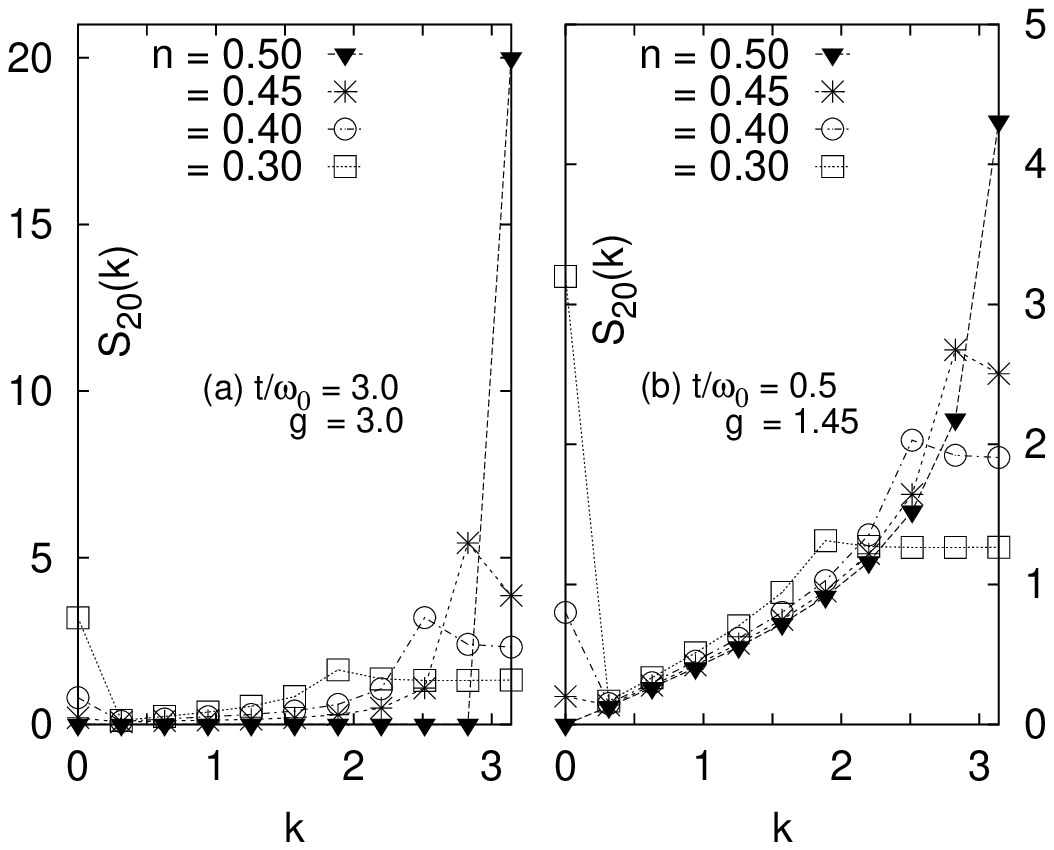}}}
\vspace*{0.5cm}
\caption[]{Structure factor plots at various values of $k$ and $n$.}
\label{scaling4}
\end{figure}

\begin{references}
   \bibitem{holstein}   
 T. Holstein, Ann. Phys. (N.Y.) {\bf 8}, 343 (1959).

  \bibitem{alex}   
 A. S. Alexandrov and N. Mott, {\it Polarons and Bipolarons} 
(World Scientific, Singapore, 1995).

  \bibitem{alex2}   
 A. S. Alexandrov and J. Ranninger, Phys. Rev. B {\bf 23}, 1796 (1981); 
 A. S. Alexandrov, J. Ranninger, and S. Robaszkiewicz,
Phys. Rev. B {\bf 33}, 4526 (1986).

  \bibitem{alex3}   
The many-polaron problem was studied for the spinless Frohlich model
 (within a {\it first order perturbation} treatment)
by  A. S. Alexandrov and P. E. Kornilovitch,
J. Phys. Condens. Matter {\bf 14}, 5337 (2002).
These authors examine the case of
 infinite-range electron-phonon coupling with zero on-site interaction
where as we deal with
the complimentary on-site interaction only case.

   \bibitem{cnr}   
For a review see {\it Colossal Magnetoresistance, Charge Ordering, 
and Related Properties of Manganese Oxides}, edited by C.N.R. Rao
 and B. Raveau (World Scientific, Singapore, 1998).

   \bibitem{dattays} These results will be reported elsewhere
by the authors (unpublished).


 \bibitem{yamaji}   
T. Ishiguro, K. Yamaji, and G. Saito,
{\it  Organic Superconductors} (Springer, Berlin, 1998); N. Tsuda, K. Nasu, 
A. Yanase, K. Siratori, {\it Electronic Conduction in Oxides} 
(Springer-Verlag, Berlin 1990).

   \bibitem{lang}   
 I.G. Lang and Yu.A. Firsov, Zh. Eksp. Teor. Fiz.  
{\bf 43}, 1843 (1962) [Sov. Phys. JETP {\bf 16}, 1301 (1962)].

   \bibitem{ys}   
 S. Yarlagadda, 
   Phys. Rev. B {\bf 62}, 14828 (2000).

   \bibitem{mars}   
 F. Marsiglio, 
   Physica C {\bf 244},                      
   21 (1995).

   \bibitem{stephan}   
 W. Stephan, 
   Phys. Rev. B {\bf 54},                      
   8981 (1996).

   \bibitem{fradkin}   
 J. E. Hirsch and E. Fradkin,
   Phys. Rev. B {\bf 27},                      
   4302 (1983).

   \bibitem{ibose}   
For a review see I. Bose, in 
{\it Field Theories in Condensed Matter Physics},
 edited by Sumathi Rao
(Institute of Physics Publishing, Bristol, 2002).

   \bibitem{gagli}   
 E. R. Gagliano, E. Dagotto, A. Moreo, and F. C. Alcaraz,
   Phys. Rev. B {\bf 34},                      
   1677 (1986); {\bf 35}, 5297 (1987).


   \bibitem{haldane}   
 F. D. M. Haldane, 
   Phys. Rev. Lett. {\bf 45},                      
   1358 (1980).

   \bibitem{luther}   
 A. Luther and I. Peschel, 
   Phys. Rev. B {\bf 12},                      
   3908 (1975).

   \bibitem{fowler}   
 M. Fowler, 
   J. Phys. C {\bf 13},                      
   1459 (1980).
  
   \bibitem{lieb}   
 E. Lieb, T. Schultz, and D. Mattis, 
   Ann. Phys. (N.Y.) {\bf 16},                      
   407 (1961).

  \bibitem{hamer}
 R. J. Bursill, R. H. McKenzie, and C. J. Hamer,
 Phys. Rev. Lett. {\bf 80}, 5607 (1998).

  \bibitem{sdys}
Infinite dimensional case  will be analyzed elsewhere by
the authors (unpublished).

   \bibitem{alex4}   
For a study of the energy dispersion of a single small polaron 
in two-dimensions 
by resummed strong-coupling perturbation theory see Ref. \cite{stephan}
and by  exact quantum Monte Carlo calculations 
see P. E. Kornilovitch and A. S. Alexandrov
 Phys. Rev. B {\bf 70}, 224511 (2004).
\end{references}
 \end{document}